\documentclass[preprint,prd,aps,showpacs,nofootinbib]{revtex4}

\newcommand{\be}{\begin{equation}}
\newcommand{\ene}{\end{equation}}
\newcommand{\bea}{\begin{eqnarray}}
\newcommand{\enea}{\end{eqnarray}}

\begin{document}

\title {\large LMA Solution to the Solar Neutrino Problem and \\
a Phenomenological Charged Lepton Mass Matrix}

\author{ Xiao-Jun Bi }
\affiliation{ Department of Physics, Tsinghua University, Beijing 100084,
People's Republic of China}
\email[Email: ]{bixj@mail.tsinghua.edu.cn}

\author{ Yuan-Ben Dai }
\affiliation{ Institute of Theoretical Physics,
 Academia Sinica, P.O. Box 2735, Beijing 100080, People's Republic of China }
\email[Email: ]{dyb@itp.ac.cn}

\date{\today}
\begin{abstract}                                                            %

We propose a phenomenological form of the charged lepton mass
matrix, which extends the idea of ``lopsided'' mass matrix in the
literature. The features of the form are that both the 2-3 and 1-3
elements of the charged lepton mass matrix are of order 1 and that
the small elements have a new structure . This form leads to an
interesting result that both large atmospheric and solar neutrino
mixing can be accounted for by the matrix.
 Another
interesting result of this mass matrix is that it leads to very
small 1-3 mixing in the lepton sector and can suppress the
branching ratio of $\mu\to e\gamma$ under the present experimental
limit in the supersymmetric see-saw case.

\end{abstract}

\pacs{14.60.Pq}
\preprint{TUHEP-TH-02136}

\maketitle


The discovery of neutrino oscillation has been one of the most exciting
experimental results in the last few years\cite{first}.
As the experimental data accumulating in the Super-Kamiokande collaboration
\cite{new,sma} and more results published by
SNO\cite{sno}, K2K\cite{k2k} and CHOOZ\cite{chooz} experiments,
the parameter space for the neutrino masses and mixing
is narrowing down considerably. The recent analyses show that the
atmospheric neutrino oscillation favors the $\nu_\mu-\nu_\tau$
process with a nearly maximal mixing angle $\sin^22\theta_{atm}\ge
0.87$ and the mass squared difference $1.5\times 10^{-3}
eV^2\le\Delta m^2_{atm}\le 4.8\times 10^{-3} eV^2$ at the 99\%
confidence level\cite{atm}. Among the four solutions for the solar
neutrino deficits, the large mixing angle MSW (LMA) solution is
most favored, followed by the LOW and VAC solutions\cite{lma}. The
small mixing angle (SMA) solution is ruled out at the 2$\sigma$
level\cite{sma,lma}. The parameter space for the LMA solution is
$0.2\le\tan^2\theta_{sol}\le 0.75$ and $2\times 10^{-5}
eV^2\le\Delta m^2_{sol}\le 4\times 10^{-4} eV^2$ at the 3$\sigma$
confidence level, with the best fit values
$\tan^2\theta_{sol}=0.37$ and $\Delta m^2_{sol}=3.7\times 10^{-5}
eV^2$\cite{lma}.

On the theoretical side, hundreds of models for neutrino masses
and mixings have been published during the last few years.
However, a survey of these models shows that most of them yield
the SMA or VAC solution for the solar neutrino problem\cite{barr},
no matter whether the models produce the tiny neutrino masses
directly at low energy or by the see-saw mechanism. In the
following we will give a brief comment on the neutrino models in
the literature and point out the difficulties in obtaining the LMA
solution for the solar neutrino problem, and then present our
model.

As we know, the neutrino mixing, which is described by the MNS
matrix\cite{mns}, is actually the mismatch between the two bases
in which the mass matrix of the charged leptons or that of the
active neutrinos is diagonal. The neutrino models are thus
generally divided into two classes according to the origin of the
large mixing angles: those from the charged lepton mass matrix
$M_L$ or from the neutrino Majorana mass matrix $M_\nu$. For the
models which generate the large mixing angles in $M_\nu$, it is
usually difficult to reconcile the large mixing angles and a small
mass squared splitting ratio $r\equiv \Delta m_{sol}^2/\Delta
m_{atm}^2 \approx 1.4\times 10^{-2}$, which generally means a
hierarchical structure of $M_\nu$. The LMA solution in this kind
of models is usually achieved by fine tuning the parameters in
non-see-saw case or by choosing a special hierarchy of the right
handed neutrino spectrum in the see-saw case\cite{barr}.

For models in which the large mixing angles coming from the
charged lepton mass matrix the situation is better, since the
large mixing and the small mass ratio come from different
matrices. These models are either based on the beautiful idea of
``flavor democracy''\cite{demo} or on the idea of ``lopsided''
form\cite{albright,lopsid} of the charged lepton mass matrix.
However, having large mixing angles in $M_L$ raises the question
why large CKM angles do not arise from the diagonalization of the
quark mass matrices $M_D$ and $M_U$. Especially, in grand-unified
models the Dirac mass matrices $M_D$, $M_U$ and $M_L$ are
generally closely related.

The answer to this question in the ``flavor democracy''
models is that the CKM angles are small by a cancellation caused
by an approximate symmetry between $M_U$ and $M_D$.
This kind of models can predict a bi-maximal form of the MNS
matrix by diagonalizing the charged lepton mass matrix and assuming
the neutrino mass matrix diagonal at the same time. However,
this kind of models tend to give a too large solar neutrino mixing
angle near maximality, which is at most marginally consistent with
the LMA parameter space.

The ``lopsided'' form of the charged lepton mass matrix is another
elegant idea to give large lepton mixing while keeping small quark
mixing at the same time. It is usually realized in the context of
grand unified models, especially if an SU(5) symmetry plays a role
in the form of the fermion mass matrices. In a SU(5) grand unified
model, the left handed charged leptons are in the same multiplets
as the CP conjugates of the right handed down quarks. Since the
large mixing angle $\theta_{atm}$ is attributed to the mixing of
the left-handed leptons in such models, it would generally be
related by $SU(5)$ to a large mixing angle for the right-handed
down quarks, which is not observable. On the other hand, the small
CKM angles are related by $SU(5)$ to small mixings of the
right-handed leptons, which are irrelevant to neutrino oscillation
phenomena.

However, in most published models the ``lopsided'' form only
accounts for the largeness of $\theta_{atm}$ and a SMA or VAC
solution for the solar neutrino problem is usually predicted.
To obtain a LMA solution in such models requires a  right handed
neutrino mass matrix with a complicate form and fine tuning of the
parameters to some extent\cite{albright,lopsid}.

One exception is given by K. S. Babu and S. M. Barr,
where  both the large solar and
atmospheric neutrino mixing angles can be accounted for by diagonalizing
the charged lepton mass matrix\cite{babu}.
In their model, the charged lepton mass matrix has the following
``lopsided'' form
\footnote{Here we use the
convention that a left-handed doublet multiplies the Yukawa coupling matrix
from the left side while a right-handed singlet multiplies the matrix from
the right side.}:
\be
\label{form1}
M_L=\left( \begin{array}{ccc} 0 & 0 & \rho ' \\ 0 & \delta & \rho \\ \delta' & \epsilon & 1
\end{array}\right) m_D \ ,
\ene with $\rho'\sim \rho \sim 1$, $\delta\sim \delta'\ll \epsilon \ll 1$.
The parameters are  determined by fitting the quark and lepton spectra and
mixing angles. The unique nontrivial prediction of the model is
$U_{e3}$ element in the MNS matrix by diagonalizing $M_L$,
which is $|U_{e3}|\simeq 0.05$. This
prediction will provide a test of this model.

However, most ``lopsided'' models predict a similar result of
$U_{e3}\simeq 0.05$. Thus a measurement of $U_{e3}\simeq 0.05$
will not discriminate this model from others. Furthermore, the
lepton flavor violating process $\mu\to e\gamma$ gives very strong
constraints on the element $U_{e3}$\cite{bi,sato}. It is pointed
out in \cite{bi} that the ``lopsided'' models, which always give a
large $\mu-\tau$ mixing and a typical value of $U_{e3}\simeq 0.05$,
may lead to the branching ratio of $\mu\to e\gamma$ exceeding the
present experimental upper limit in the supersymmetric see-saw
case. This will be explained below. Actually, most ``lopsided''
models are realized in supersymmetric unified SO(10) models and
the neutrino spectrum and mixing are given by see-saw mechanism.
Thus, it seems that all these models meet the same difficulty in
predicting the branching ratio of $\mu\to e\gamma$.

We have found a form of the charged lepton mass matrix
 independently when we tried to construct a
``lopsided'' model which can account for the LMA solution of the
solar neutrino problem and simultaneously avoid the above
difficulty on $\mu\to e\gamma$ decay. Our model is similar to the
form (\ref{form1}) in the large elements in the third column but
different in the structure of the first two columns. It gives a
completely different prediction of $U_{e3}$ and the branching
ratio of $\mu\to e\gamma$ can be below the present experimental
limit. In the following we shall first give our model and its
features and then explain the smallness for $U_{e3}$. In the last
we shall discuss briefly its implication on the lepton flavor
violating process $\mu\to e\gamma$ in supersymmetric see-saw case.

The aim of the work is to present a phenomenological model
to account for
the neutrino oscillation data and avoid the $\mu\to e\gamma$ difficulty.
The mass matrices of the charged lepton and neutrino are
assumed to have the following form:
\be
\label{form}
M_L=\left( \begin{array}{ccc} 0 & \delta & \sigma \\ -\delta & 0 & 1-\epsilon\\
0 & \epsilon & 1 \end{array}\right) m,\ \
\ene
\be
\label{form2}
M_\nu=\left(
\begin{array}{ccc} m_1 & 0 & 0 \\ 0 & m_2 & 0\\ 0 & 0 & m_3
\end{array}\right)\ ,
\ene
with $\sigma\sim \mathcal{O}(1)$, $\epsilon\ll 1, \delta\ll
\epsilon$.
Similar to form (\ref{form1}) the model extends the usual
``lopsided''  form of the charged lepton
mass matrix between
the second and the third generations to include the first
generation.
For simplicity we have assumed that $M_\nu$ is diagonal here. The
main feature of the model will not be changed if mixing in
$M_\nu$ is small. Taking the values of the parameters \be
\label{para} \delta=0.00079,\ \epsilon=0.12,\ \text{and}\ \
\sigma=0.55
\ene
we can  obtain the correct mass ratios
$m_e/m_\mu$, $m_\mu/m_\tau$ and the MNS matrix
\be
\label{vmns}
V_{MNS}=\left(\begin{array}{ccc} 0.851 & -0.525 & -0.0056 \\
    0.362 & 0.595 & -0.718 \\ 0.380 & 0.609 & 0.696 \end{array} \right)\ .
\ene This model then predicts the neutrino mixing parameters as
\be \sin^22\theta_{atm}=0.999, \tan^2\theta_{sol}=0.38\
\text{and}\ U_{e3}=-0.0056 .
\ene

The notable feature of form (\ref{form}) compared with the usual
``lopsided'' models is the order $1$ element
$\sigma$. We can see from Eq. (\ref{vmns}) that by choosing a
large $(2,3)$ and $(1,3)$ elements in $M_L$ we get two large
mixing angles, which are corresponding to the maximal mixing
of the atmospheric neutrinos and the large solar neutrino mixing
in the LMA solution,
by diagonalizing the charged lepton mass matrix.  In this
phenomenological model we separate the large
angles $\theta_{atm}$ and $\theta_{sol}$ from the small mass ratio
$\Delta m_{sol}^2/\Delta m_{atm}^2$ completely. Thus it is very
easy to reconcile the large mixing angles and the neutrino
spectrum of the LMA solution without any fine tuning.

Unlike the ``flavor democracy'' models, which usually predict a
bi-maximal MNS matrix, our model gives the best fit value of $\theta_{sol}$
for the LMA solution. It is unlike the usual ``lopsided'' models
either, which predict a small lepton mixing angle between the
first and the second generations with a typical value of
$\sqrt{m_e/m_\mu}\approx 0.07$\cite{albright,lopsid}. This phenomenological
form may provide a new possibility of model-building for the fermion masses
and mixing.

Besides $\theta_{atm}\approx \pi/4$ the prediction of
$U_{e3}=-0.0056$ is quite non-trivial, since all the parameters
are fixed by the lepton mass ratios and the solar neutrino mixing
angle. This prediction is different from most other ``lopsided''
models, including that given in Ref. \cite{babu}.
It thus provides a test of our model.

The smallness of $U_{e3}$ can be explained as following.
The charged lepton mass matrix $M_L$ can be diagonalized by
a bi-unitary rotation
\be
U_L^\dagger M_L U_R = \left( \begin{array}{ccc} m_e & & \\ & m_\mu & \\
&&m_\tau \end{array}\right)\ .
\ene
The MNS matrix is defined by $U_{MNS}=U_L^\dagger U_\nu$, where $U_\nu$
is the unitary matrix diagonalizing the neutrino mass matrix
and $U_\nu = I$ here. Thus $U_{MNS}$ is given by $U_L$. $U_L$ can
be approximately obtained by rotating the left-handed charged leptons
so that the right-upper triangle of $M_L$ in (\ref{form}) becomes zeros.
This can be achieved by three steps. First we rotate the
first two generation charged leptons by a matrix
\be
V_{12} = \left( \begin{array}{ccc} c_{12} & s_{12} & 0 \\
-s_{12} & c_{12} & 0 \\ 0 & 0 &1 \end{array}\right) \ ,
\ene
with $ c_{12}=\cos\theta_{12}$, $ s_{12}=\sin\theta_{12}$ and
$\tan\theta_{12}=\sigma/(1-\epsilon)$. Then the $1-3$ element of $M_L$
becomes zero and the $2-3$ element becomes
$\sigma'= \sqrt{\sigma^2+(1-\epsilon)^2}$.
The second step is a similar rotation between the second and the third
generations, $V_{23}$, with $\tan\theta_{23}=\sigma'$,
which makes the $2-3$ element of $M_L$ zero. The final step is a small
rotation $V'_{12}$ which makes the $1-2$ element zero. The rotation
angle is $\theta'_{12}\cong -\frac{c_{12}\delta}{s_{23}\epsilon}$.
We then get the unitary matrix
\be
\label{ul}
U_L^\dagger = {V'}_{12}^\dagger V_{23}^\dagger V_{12}^\dagger\ ,
\ene
from which we have that $\theta_{atm}\cong -\theta_{23}$,
$\theta_{sol}\cong -\theta_{12}$ and $U_{e3}\cong \theta'_{12} \sin\theta_{23}$.
Using the expressions of the diagonal elements of the resulting
triangular matrix  $U_{e3}$ can be approximately expressed by mass ratio and
physical mixing angles as 
\be 
\label{ue3} 
U_{e3}\cong\frac{m_e}{m_\mu}\cdot U_{\mu 3}/\tan\theta_{sol}\ . 
\ene 
On the other hand, the $U_{e3}$ from (\ref{form1}) can be expressed by
\be 
\label{ue31} U_{e3}\cong -\sqrt{\frac{m_e}{m_\mu}}\cdot U_{\mu 3}\ .
\ene 
From the Eqs. (\ref{ue3}) and (\ref{ue31}) we can see
that, for large $\theta_{sol}$, $U_{e3}$ in our model is
suppressed compared to that in Ref. \cite{babu} due
to the smallness of ${m_e}/{m_\mu}$. The difference in the
dependence of $U_{e3}$ on the mass ratio of $m_e/m_\mu$ 
originates from the different structures in
 $M_L$ in these two models.

A $U_{e3}\sim 0.05$ is at the edge of the parameter space
measurable in the next generation long baseline neutrino
experiments, which may improve the present sensitivity to
$\sin^22\theta_{13}$ by an order of magnitude, probing
$\sin^22\theta_{13}$ at the level of 0.01\cite{lbl}. To
discriminate the two kinds of models, an entry-level neutrino
factory is necessary, which can probe $\sin^22\theta_{13}$ down to
$10^{-3}$\cite{nfac1}. If no $\nu_\mu-\nu_e$ signal is observed
even at this level, most of the ``lopsided'' models may then be
ruled out. If  $U_{e3}$ is indeed as small as that in our model,
the high-performance neutrino factory, which can probe
$\sin^22\theta_{13}$ down to the order of $10^{-4}$, will be
necessary to measure its value\cite{nfac2}.

As for the quark sector, we will not discuss it in detail here. We
only want to point out that a large $(1,3)$ element in $M_L$ will
not lead to a large quark mixing if the relation $M_D\sim M_L^T$
is satisfied. Unlike the large elements in the mass matrices, the
relations between the small  matrix elements in the quark sector
and those in the lepton sector depend on the details of the
grand-unified models used. The simplest possible form of the down
quark mass matrix is obtained by first adding a small $(3,1)$ element
and a coefficient $-\frac{1}{3}$ in front of $\epsilon$ in $M_L$,
and then transposing it. Then $M_D$ can produce acceptable down
quark spectrum and mixing angles.

Finally, we discuss the case that the neutrino mass matrix is
generated by the see-saw mechanism, \be \label{ss}
M_\nu=-M_NM_R^{-1}M_N^T\ , \ene where $M_N$ is the neutrino Dirac
mass matrix.  In this case we can modify our assumption to that
the charged lepton mass matrix of the form in Eq. (\ref{form}) is
given in the basis where the Dirac neutrino mass matrix $M_N$ is
diagonal. Generally the right-handed Majorana mass matrix $M_R$
may be non-diagonal. Then $M_\nu$ in Eq. (\ref{ss}) is not
diagonal and deviates from Eq. (\ref{form2}). If mixing in $M_R$
is small, this deviation is small and the main feature of our
model described above remains intact. Another possibility is that
$M_N$ is  hierarchical and this structure  transfer to $M_\nu$
through the see-saw mechanism\cite{barr}. Taking hierarchy of
$M_N$ similar to that of up quarks as suggested in SO(10) grand
unified models and assuming no large hierarchy among the elements
of $M_R^{-1}$, we find that the mixing in $M_\nu$ is tiny, which
almost do not change the values of $V_{MNS}$ in Eq. (\ref{vmns}).

The main virtue of the assumption that $M_L$ has the form
(\ref{form}) in the basis where $M_N$ is diagonal is that it can
avoid the contradiction with the $\mu\to e\gamma$ experiments when
the ``lopsided'' form is realized in the supersymmetric see-saw
case. In this case and if supersymmetry breaking is mediated by
gravity the soft SUSY breaking terms will introduce new lepton
flavor violating sources. The mechanism is that there are two
Yukawa coupling matrices $Y_L$ and $Y_N$ in the lepton sector for
the charged leptons and neutrinos at the energy scale above the
right-handed neutrino masses. They are proportional to $M_L$ and
$M_N$ respectively. These two Yukawa coupling matrices can not be
diagonalized simultaneously and will lead to lepton flavor mixing,
in analogy to the flavor mixing in the quark sector. This lepton
flavor mixing can transfer to the scalar lepton sector through
quantum effects. The masses of sleptons are usually assumed to be
 universal at the SUSY breaking scale to
avoid the low energy flavor problems. The flavor mixing effects in
the scalar lepton sector in the low energy region can then be
given approximately by 
\be 
\left(\delta
m_{\tilde{L}}^2\right)_{ij} \approx\frac{1}{8\pi^2}
V_{i3}V^*_{j3}\cdot Y_{N_3}^2(3+a^2)m_0^2\log\frac{M_{GUT}}{{M_R}}
\ \ , 
\ene 
where $(\delta m_{\tilde{L}}^2)_{ij}$ is the
non-diagonal terms of the scalar lepton mass matrix, representing
the lepton flavor mixing between the $i$-th and the $j$-th
generations. $Y_{N_3}$ is the third generation Yukawa coupling of
neutrinos, which dominates over the first two generations. $V$ is
the mixing matrix in the lepton sector. The loop effect with
internal SUSY particles is much more effective than that with internal
neutrinos in producing the $\mu\to e\gamma$ process so that the
rate of this process is determined by $(\delta
m_{\tilde{L}}^2)_{12}$, which is proportional to $V_{23}V^*_{13}$.
Since there is a very high experimental sensitivity to the process
 $\mu\to e\gamma$ ($Br(\mu\to e\gamma)<1.2\times 10^{-11}$\cite{exp}),
and $V_{23}$ is large (around $1/\sqrt{2}$) in ``lopsided'' models,
$V_{13}$ is thus very strongly constrained.

In the usual ``lopsided'' models, with the typical value of
$U_{e3}\sim 0.05$, the branching ratio of $\mu\to e\gamma$ is then
given by\cite{bi} 
\be 
Br(\mu\to e\gamma)\approx  C\cdot 10^{-7}
\left( \frac{100\, GeV}{m_s} \right)^4 \left( \frac{\tan\beta}{10}
\right)^2\ , 
\ene 
where $C$ is around $1\sim 10$. Thus, if the common SUSY
particle mass scale, $m_s$, is $\lesssim 1 TeV$ and $\tan\beta$ is not too small,
then the branching ratio should be above the present experimental
limit. This fact has been noticed by a few authors\cite{bi, sato}.
However, in our model, by the assumption that $M_L$ has the form
 (\ref{form}) in the basis where $M_N$ is diagonal, $V_{13}$ 
can be as small as $-0.0056$. Since the rate for the process
$\mu\to e\gamma$ is approximately proportional to $V_{13}^2$  it
is greatly suppressed compared to it's value in the usual
``lopsided''models. Our model predicts a branching ratio of
$\mu\to e\gamma$ not too much smaller than the present
experimental limits and easily to be detected in the future
experiments\cite{nexp}. This is an important difference of our
model from the model of Ref. \cite{babu}. Like the usual
``lopsided'' models, Ref. \cite{babu} predicts  a $\mu\to e\gamma$
branching ratio already exceeding the present experimental limit.

The process $\tau\to\mu\gamma$ is another
promising process to determine whether there is a large mixing
between the second and the third generations in the charged lepton
sector, since its rate is proportional to $|V^*_{23}V_{33}|^2$.
If both processes are found in the future experiments in
the range of $10^{-6} > Br(\tau\to\mu\gamma) > 10^{-9}$ and
$10^{-11} > Br(\mu\to e\gamma)
> 10^{-14}$\cite{exp,nexp}, our model, with a large $2-3$ mixing while
very small $1-3$ mixing, will be a very attractive
candidate to accommodate the observations.

In summary, we have constructed a phenomenological form of the
charged lepton mass matrix, which provides a new possibility in
model construction. This form can produce the large solar neutrino
mixing angle and the maximal atmospheric neutrino mixing angle
simultaneously without any fine tuning. With such a charged lepton
mass matrix it is very easy to build a neutrino model to explain
the observed neutrino oscillation experiments. The unique
prediction of this model is that $U_{e3}=-0.0056$, which is
different from predictions of all other ``lopsided'' models . This
prediction can be tested in neutrino factory.  This model has
another virtue that it predicts the branching ratio of $\mu\to
e\gamma$ below  the present experimental limit but may be not too
much smaller than it in the SUSY-GUT see-saw case. If both the
$\tau\to\mu\gamma$ and $\mu\to e\gamma$ processes are detected in
the near future experiments, our model will be an attractive
candidate to explain the experimental results.

\begin{acknowledgments}
This work is supported by the National Natural Science Foundation
of China under the grand No. 10105004 and No. 19835060.
\end{acknowledgments}

\end{document}